\documentclass[12pt,onecolumn,draftclsnofoot]{IEEEtran}
\usepackage[utf8]{inputenc}
\usepackage{graphicx}
\usepackage[tbtags]{amsmath}
\usepackage{amssymb}
\usepackage{cite}
\usepackage{lettrine}
\usepackage{tkz-graph}
\usepackage{algorithm}
\usepackage[noend]{algpseudocode}
\usepackage{graphicx}
\usepackage{subcaption}
\usepackage{lipsum}
\usepackage{enumitem}
\usepackage{mathtools}
\usepackage{footnote}
\usepackage{tabularx}

\begin{document}
\title{Fading Improves Connectivity in Vehicular Ad-hoc Networks}
\author{Samar Elaraby,~\IEEEmembership{Student Member,~IEEE} and Sherif M. Abuelenin

\thanks{S. Elaraby (samar.elaraby@eng.psu.edu.eg) 
and S. M. Abuelenin (s.abuelenin@eng.psu.edu.eg) are with the Department of Electrical Engineering, Faculty of Engineering, Port Said University, Port Said 42526, Egypt.}
\thanks{Manuscript received XXX, XX, 2020; revised XXX, XX, 2020.}}

\markboth{Preprint}
{}
\maketitle
\begin{abstract}
	Connectivity analysis is a crucial metric for network performance in vehicular ad-hoc networks (VANETs). Although VANET connectivity has been intensively studied and investigated under no-fading channel models for their simplicity, these models do not represent real-world scenarios that suffer channel impairments. The connectivity probability in a multipath propagation environment is too challenging to be caught by a closed formula due to the emerging complexity associated with the randomness in a fading channel. This leads to contradicting statements about the impact of fading on VANET connectivity. In this paper, we numerically estimate the connectivity probability using graph-based Monte-Carlo simulations aiming for better understanding of the connectivity in fading channels. The results show that Rayleigh-fading channels reinforce the connectivity compared to no-fading models at the same level of transmitting power and vehicle densities. While these findings may seem counterintuitive, they agree with similar behavior that was reported earlier in other ad-hoc networks. Using simulations and stochastic analysis, we thoroughly investigate this effect and provide an intuitive interpretation of the positive impact of fading on connectivity. 
\end{abstract}
\section{Introduction}

\lettrine{C}{onnectivity} of vehicular ad-hoc networks (VANETs) has been well investigated in the literature as a crucial metric for network performance. VANETs applications range from safety and traffic alert dissemination to dynamic routing planning to entertainment, gaming and content sharing to Internet of vehicles (IoV) and serve as a technology accelerator for autonomous vehicles \cite{Ucar2016, Darwish2018, Toor2008, Yang2014a, Kaiwartya2016, Maalej2018}. 
However, there are still gaps in our understanding of VANETs connectivity in real propagation environments.

In VANETs, vehicles communicate with each other in one- or multi-hop routes without the need of an infrastructure. Therefore, VANETs represent a special class of mobile ad-hoc networks (MANETs) having their nodes confined in predetermined one-dimensional (1D) roads \cite{Toor2008}. Due to the high mobility of nodes, VANETs are distinguished from other MANETs by constantly changing topology \cite{kim2011reliable} and varying node density. Therefore, VANETs can suffer high probability of network partitioning, and no guarantee of end-to-end connectivity \cite{Toor2008, abdalla2007current}. 
The connectivity analysis is influenced by the \textit{traffic headways}, vehicle mobility and the communication environment. Traffic headway is defined as the bumper-to-bumper distance, i.e. the spacing between the front of one vehicle to the front of the next, which we refer to as its \textit{successor}. In free-flow traffic, vehicle arrivals follow a Poisson random process, and therefore, the traffic headways are widely accepted to be exponentially distributed. Other theoretical and empirical models were suggested for better headway modeling under other traffic conditions \cite{Krbalek2000, Abul-Magd2007, Nagel2010, Cheng2012, Li2017}. 

On the other hand, communication channel modeling significantly impacts connectivity analysis. Any two vehicles are said to be connected if each is located within the \textit{communication range} of the other. 
The communication range is typically defined as the distance from the transmitter within which the received signal-to-noise ratio (SNR) exceeds a certain threshold value. The shape of the corresponding coverage area changes accordingly with the wireless communication channel conditions.

In literature, several studies relied on the fixed communication range model \cite{Yang2014, Khan2017, Chen2018, Meng2018, Abuelenin2015, Abuelenin2018, Abuelenin2014, Panichpapiboon2008, Cheng2012, Ng2011, Jin2011}, also known as the unit disc model \cite{Akhtar2015, Akhtar2013, Naboulsi2017}, as the signal propagation model. In such model, the communication range is restricted to a predefined circle with the transmitting vehicle in its center. Such model neglects the effects of the communication channel impairments and greatly reduces the complexity of connectivity analysis. Besides its simplicity, the popularity of the unit disc model relates to the assumption that it conveys the average behavior of the vehicular network \cite{Naboulsi2017}. Nevertheless, using this model was criticized in literature. For example, in [23], the authors used simulations to prove that the average \textit{node degree} under the unit disc model, which is the average number of vehicles connected to one vehicle, is comparable to realistic situations only at low values of the communication range (below 150m), but largely diverges at higher values. As we discuss in the related work section, several articles have considered more realistic channel models in analyzing VANETs connectivity. 

A challenging task was to include the small-scale fading effects in connectivity analysis. In real situations, multipath fading affects the reachability of nodes and accordingly the network connectivity. Random fluctuations of the received signal strength due to the constructive and destructive interference of multipath components can cause the received signal to be stronger (weaker) at farther (nearer) locations from the transmitter, and accordingly, the concept of coverage area ceases to be applicable. To the best of our understanding, an exact expression of network connectivity probability under fading conditions cannot be reached. The works that considered fading in VANET connectivity had to rely on different assumptions and simplifications that resulted in reaching inaccurate and contradicting findings among different articles. Such findings include that, in comparison to including fading in the analysis, the unit disc approach barely impacts the global connectivity of the network, that fading would negatively affect the connectivity, and that fading could have a positive impact on the network connectivity. This is discussed in detail in the related work section.

This paper aims to provide a better understanding of the effect of fading on connectivity of VANETs. We propose using graph-based Monte-Carlo simulations to estimate the connectivity probability efficiently and accurately. The main findings presented in this paper include showing that, to the contrary of earlier conclusions, fading improves the connectivity in VANETs and that the unit disc cannot be used to accurately analyze the fading-based network topology.

\subsection{Related Work}

Different studies have considered connectivity analysis under channel impairments \cite{Miorandi2006, Ukkusuri2008, Chandrasekharamenon2012, Babu2013, C.Neelakantan2012}. 
More specifically, \cite{Miorandi2006} analyzed the effect of log-normal shadowing and Rayleigh fading on connectivity by finding an expression for the probability that any two nodes are connected. The authors assumed that this probability was equal to the probability that the distance between the two nodes is less than the communication range. We discuss the inaccuracy of this assumption in in Section \ref{V2V_Connectivity}. References \cite{Ukkusuri2008, Chandrasekharamenon2012, Babu2013, C.Neelakantan2012} followed similar methodology, but considering additional fading models. Their method described the connectivity between any two consecutive vehicles, but it becomes challenging when analyzing the whole network connectivity. Defining the network connectivity probability as the probability that each two successive nodes are connected (e.g. \cite{Babu2013}) is inaccurate, because channel impairments can cause one node to be unreachable by closer nodes while, at the same time, be reachable by a node that is farther away. 

Several articles agreed that channel impairments, especially fading, would negatively impact the connectivity of VANETs as it does for point-to-point communication systems. In addition to the deteriorated node degree reported in \cite{Akhtar2013}, the authors of \cite{ChoongHockMar2005} suggested that fading channels require higher transmission power or vehicle density to achieve the same level of connectivity as of the unit disc model (no-fading). In \cite{Miorandi2008}, it is explicitly stated that Rayleigh-fading channels have negative impacts on connectivity of ad-hoc networks, and there would be no improvement in connectivity except by means of diversity. Moreover, the connectivity probability of VANETs derived in \cite{Babu2013} shows deteriorated connectivity in fading channels compared to the unit disc channel.

On the contrary, while the authors of \cite{Bettstetter2005} were investigating the impact of shadowing on connectivity, they argued that the higher the fading variance is, the higher the connectivity probability of MANETs becomes, i.e. fading can help the network to become connected \cite{Bettstetter2005}. Also, \cite{Hekmat2006} suggested that fading increases the probability of long links yielding to improved connectivity in ad-hoc networks. Refs. \cite{Zhou2008, Georgiou2014} reported similar insights under certain conditions for static wireless ad-hoc networks (WANETs) that have their nodes uniformly distributed in a 2D space. A related remark was suggested by the authors of \cite{booth2002ad, franceschetti2005continuum} using percolation theory. The percolation threshold (the critical node degree above which an unbounded connected component exists and below which the network is certainly disconnected) serves a good indicator for connectivity \cite{haenggi2005routing}. Their theoretical work suggested that longer unreliable connections can substantially improve the connectivity of stochastic networks, even when some shorter links are lost \cite{franceschetti2005continuum}.

These findings should be thoroughly assessed before they are generalized to VANETs due to the problem restrictions. Not only do vehicles constitute 1D queues, but also, they employ different headway distributions than nodes in other networks. And the vehicle densities, considering free-flow traffic, are much lower than node densities in other ad-hoc networks. 
\subsection{Paper Organization}
First, we define the system model used to analyze VANET connectivity in Section \ref{Conn}. Then, the connectivity analysis under the unit disc assumptions is discussed in Section \ref{Model}. Graph-based simulations are introduced in Section \ref{Graph}, while in Section \ref{Rayleigh} we derive the connectivity probability of the single-link and vehicle connectivity under a Rayleigh-fading channel model. Simulations results are provided in Section \ref{Improvement} to prove the positive impact of fading on connectivity, before the conclusions are presented in Section \ref{Conclusions}.
\section{System Model and Connectivity Definitions} \label{Conn}
This section introduces the VANET highway model used in this paper along with the channel models adopted for comparisons. In addition, we introduce four different definitions of connectivity in order to fully understand the network connectivity and the proposed claims.
\subsection{Highway Free-Flow Traffic Model}
The considered VANET model is of a multi-lane segment of a highway with a length of $L$. Due to vehicle sparseness on a highway, the vehicles, with a vehicle density $\rho$, are free to choose their own speed. Thus, their movements are independent of each other. This behavior is captured by the free-flow traffic model \cite{Abul-Magd2007}. 
According to traffic theory, a sensor placed along a highway observes the arrival of vehicles as a Poisson process in free-flow traffic. Consequently, the intervehicle spacings between every two successive vehicles were proved to be i.i.d. random variables with exponential distributions \cite{Yousefi2008}. Let the intervehicle spacing between vehicle $v_{i}$ and $v_{i+1}$, where $i = 1, 2, \dots, N-1$, be a random variable $\text{X}_i$ and its probability density function (PDF) be
\begin{equation}
    f_{\text{X}_i}(x) = \rho \ e^{-\rho x}, \quad x \geq 0
    \label{PDF_One}
\end{equation}
Since the lane separation is extremely small compared to the intervehicle spacing along the road, the former is commonly neglected under the free-flow traffic conditions.
  \subsection{Channel Models}
We consider two different channel models, which are the unit disc and Rayleigh-fading model. In the unit disc model, we assume that one vehicle can transmit its packets to its neighbors within a circular disc with a fixed radius, called the communication range $r$. Within this area, the received SNR is maintained higher than a predefined threshold $\Psi$.  

Furthermore, we consider a small-scale fading model, whose received SNR can be represented by a random variable. In vehicular communications, different models, e.g., Rayleigh, Rician, Nakagami-m, or Weibull fading models, have been adopted in the literature \cite{Babu2013, Chandrasekharamenon2012, Zeng2019, Jameel2017, Renaudin2013, Molisch2009}. Less-severe fading environments with a dominant line-of-sight (LoS) are represented by a Rician model, while Rayleigh fading is adequate when it is absent. Both Nakagami-m and Weibull models are parametric and can express fading with varying severity. Under the assumption of the free-flow traffic, vehicles are wide apart leading to eliminating the LoS between them. Therefore, Rayleigh fading is the best fit for the channel \cite{Acosta-Marum2007}.

We elaborate on the details of the two models in Sections \ref{Fixed CR} and \ref{V2V_Connectivity}, respectively.
  \subsection{Connectivity Definitions}
The definition of network connectivity can vary according to applications if considered from a physical-layer perspective. The one considered in this paper copes with the extreme case, where all nodes should be connected altogether to form one cluster at any given time. In other words, there is a connection link between each pair of nodes through a single- or multi-hop route at any given time \cite{Bettstetter2002}. In broadcasting applications, for example, the delivery of broadcast messages to all the nodes on a certain road segment is then sustained with considerable delays. The connectivity analysis estimates the probability that the network is connected on the physical layer. Although the network topology would change dynamically with vehicle mobility, this paper, like \cite{Miorandi2006, Ukkusuri2008, Chandrasekharamenon2012, Babu2013, C.Neelakantan2012,  Khan2017, Meng2018, Abuelenin2015, Abuelenin2018, Abuelenin2014, Panichpapiboon2008, Cheng2012, Ng2011, Akhtar2015, Jin2011, Akhtar2013, Naboulsi2017}, considers the \textit{instantaneous connectivity}, which limits the analysis to a snapshot where all vehicles are considered fixed. The instantaneous connectivity helps understand the average behavior of the network connectivity and provides an estimate the minimum average vehicle density that maintains a certain level of connectivity.

In this paper, we claim that fading would help improve the network connectivity. In order to justify our claim, we rely on different definitions of connectivity in VANETs, summarized in Table \ref{Sym}.
\begin{table}[t]
\captionsetup{justification=centering, labelsep=newline, textfont=sc}
   \caption{Connectivity and isolation probability definitions used in the paper.} 
   \centering
   \begin{tabularx}{\textwidth}{lcX} 
      \hline
      \hline
          Parameter & Symbol  & Definition\\
      \hline
       Single-link connectivity & $P_{SL|Ray}$ ($P_{SL|UD}$)\footnotemark & 
       The probability that two nodes are connected with a direct link. Each vehicle that maintains a single link with another vehicle is called its \textit{linked neighbor}. \\
       Two-side vehicle isolation & - & The probability that a vehicle has no linked neighbors at all.\\
       One-side vehicle isolation & - & Each vehicle has neighbors both in front of and behind it (see Fig. \ref{FixedCR_Network}), which are known as \textit{forward} and \textit{backward neighbors}, respectively. One-side vehicle isolation is then the probability that a vehicle is not connected to at least either their forward or backward neighbors.\\
      Vehicle connectivity & $P_{V|Ray} (P_{V|UD})$  & The probability that a vehicle is not isolated. Thus it represents the complement event of vehicle isolation, and accordingly, has one- and two-side definitions.\\
      Network connectivity  & $P_{c|Ray} (P_{c|UD})$ & The probability that all the nodes form one unpartitioned, connected network.\\
       \hline
       \hline
   \end{tabularx}
   \label{Sym}
\end{table} \footnotetext[1]{The subscripts $Ray$ and $UD$ refers to the Rayleigh fading and unit disc models, respectively.}
%
In the following sections, we compare those parameters in both the unit disc and Rayleigh-fading models for better understanding the effects of fading on connectivity.
\section{Connectivity Under the Unit Disc Model} \label{Model}
The unit disc model neglects the randomness of the communication channel. The signals are only exposed to the path loss, described in \eqref{free-space}. Therefore, the vehicles can transmit to its neighbors within a disc area. However, the unit disc model is widely used in the literature for its simplicity. In the following, we derive the connectivity probability under its assumptions for the comparisons with the Rayleigh-fading model.
\subsection{Single-Link Connectivity}\label{Fixed CR}
The instantaneous single-link connectivity between any two vehicles depends on the transmitting power, vehicle density, and communication channel \cite{Bettstetter2005}. We assume that the power transmitted from all vehicles is identical. Following the free-space propagation model, the received SNR at distance $d$ can be defined as
\begin{equation}
    \gamma =  \dfrac{\beta P_T}{d^{\alpha} P_{noise}}
    \label{free-space}
\end{equation}
where $\beta = \zeta g$, $\zeta$ is the path loss at a reference distance of $1 \ m$, and $g$ is the total antenna gain. $P_T$ is the transmitted power, $\alpha$ is the path-loss exponent (PLE), and $P_{noise}$ is the noise power. The communication range $r$ can be determined from \eqref{free-space} as the distance $d$ that maintains a certain SNR threshold $\Psi$, i.e., 
\begin{equation}
	r = \left(\frac{\beta P_T}{\Psi P_{noise}}\right)^{1/\alpha}
\end{equation}
Therefore, a vehicle is a linked neighbor of another if it receives the latter's transmissions with an SNR higher than $\Psi$. Another approach is to check whether  it is located within a distance $r$ from the transmitter. The two approaches give identical results in the unit disc model. 

In the unit disc model, a vehicle that is connected to its second nearest neighbor definitely has a link to its first successive neighbor, the connections to far neighbors contain redundant information. Therefore, direct links to farther neighbors are omitted in the probabilistic analysis. Any two successive vehicles are connected if their intervehicle spacing is smaller than the fixed communication range. Thus, the single-link probability becomes
\begin{equation}
        P_{SL}^{(1)} = \mathbb{P}(\text{X}_i \leq r)
        = F_{\text{X}_i}(r)
        = 1 - e^{- \rho r}
\label{exponential_Conn}
\end{equation}
where the superscript of $P_{SL}^{(1)}$ refers to the first successive neighbor restriction, and  $F_{\text{X}_i}(r)$ is the cumulative distribution function (CDF) of the intervehicle spacing \cite{Panichpapiboon2008}.
\subsection{VANET Connectivity Under the Unit Disc}\label{UDConn}
The network connectivity, accordingly, is the probability that each vehicle is connected to its successor. For a network of $N$ vehicles, the network connectivity requires the existence of $N-1$ links, each of which connects two different adjacent vehicles. Hence, the network connectivity probability can be evaluated as discussed in \cite{Panichpapiboon2008} by
\begin{equation}
    P_{c|UD} = \prod_{i=1}^{N-1} \mathbb{P}(\text{X}_i \leq r) = \left(1 - e^{- \rho r}\right)^{N-1}
    \label{Eq:FixedCR_Connectivity}
\end{equation}
Since $r$ is assumed to be fixed, the VANET connectivity depends on both vehicle density and the total number of vehicles. It can be inferred that, for the same road segment, the higher the average vehicle density, the higher the connectivity probability we can achieve.
\section{Graph-based Simulations for Connectivity Estimation} \label{Graph}
A vehicular network can be represented as a graph $\mathcal{G}(\mathcal{V}, \mathcal{E})$, where the vertices set $\mathcal{V}$ contains the vehicles and the edges set $\mathcal{E}$ has the connections between the vehicles. In our analysis, we represent VANETs under the unit-disc and fading models with unweighted, undirected graphs, which is also exploited by \cite{Bettstetter2002, Naboulsi2017, Han2006, Hoque2014, Ukkusuri2008} among others. The choice of undirected graphs for VANETs follows a channel reciprocity assumption. 

Graphs can be represented mathematically by two different types of matrices. First is the adjacency matrix that represents the graph structure, i.e., its vertices and edges. The adjacency matrix $\mathbf{A}$ of an unweighted, undirected graph is a symmetric matrix whose elements $A_{ij}$ and $A_{ji}$ equal one if there exists an edge between the nodes $v_i$ and $v_j$ and zero otherwise. Second is the Laplacian matrix, $\mathbf{L} = \mathbf{D} - \mathbf{A}$, which contains both of edge information in its off-diagonal elements and node degrees in its diagonal. The matrix $\mathbf{D}$ is the degree matrix.
Moreover, the Laplacian matrix differentiates between \textit{partitioned} and connected graphs. The number of partitions in one graph can be evaluated using the eigenvalues of the Laplacian matrix. The number of zero eigenvalues is itself the number of the partitions in the graph represented by that Laplacian matrix \cite{Stankovic2019}. Since all the eigenvalues of a real Laplacian matrix are positive, the second smallest eigenvalue, called \textit{algebraic connectivity}, determines the graph connectivity \cite{Fiedler1973}.
\begin{algorithm}[!t] 
 \caption{A pseudocode of Monte-Carlo simulations}
 \begin{algorithmic}[1]
\State Set the value of the SNR threshold value $\Psi$
\State Set a matrix $\mathbf{W}$ as a zero matrix of size $N \times N$
\State Set $counter = 0$
\For{every iteration}
\State Set a vector $\mathbf{y}$ with $N-1$ random numbers that are exponentially distributed with a mean of $\frac{1}{\rho}$
\State Calculate the upper triangular part of $\mathbf{W}$ such that {$\forall$ $j > i$} $W_{ij} \coloneqq  \sum_{m = i}^{j-1} y_{m}$ 
\State Calculate the lower triangular part of $\mathbf{W}$ such that $W_{ji} \coloneqq W_{ij}$
\State Set matrix $\mathbf{P}$ for the received SNR with $P_{ij} \coloneqq \frac{\beta P_T}{W_{ij}^{\alpha} P_{noise}}$ 
\State $\mathbf{A} \coloneqq \mathbf{P} \geq \Psi$
\State Set the diagonal degree matrix $\mathbf{D}$ with its diagonal elements $d_{ii} \coloneqq  \sum_{j=1}^{N} A_{ij}$
\State $\mathbf{L} \coloneqq \mathbf{D} - \mathbf{A}$
\State Set $\lambda_2$ to the second-smallest value of eign$(\mathbf{L})$
\If {$\lambda_2 \neq 0$}
\State $counter \coloneqq  counter + 1$
\EndIf
\EndFor
\State Calculate $P_{c} \coloneqq  counter / \#iterations$
\end{algorithmic}
\label{algo}
\end{algorithm}

Based on the graph algebraic connectivity, we estimate the connectivity probability of highway VANETs using Monte-Carlo simulations. A pseudocode of the procedure is presented in Algorithm \ref{algo} under the unit disc model. First, a weighted matrix $\mathbf{W}$ is generated to represent the exponentially-distributed intervehicle spacings. In order to examine the connectivity between any two nodes, we rely on the received SNR and check whether it exceeds the threshold $\Psi$. We avoid using the communication range approach because of its inadequency when considering Rayleigh-fading channels as we show in Section \ref{V2V_Connectivity}. So we construct the matrix $\mathbf{P}$ using \eqref{free-space}. Then, the adjacency matrix $\mathbf{A}$ is computed by comparing each element of $\mathbf{P}$ to $\Psi$. Next, the second-smallest eigenvalue is determined from the Laplacian matrix $\mathbf{L}$. If the second-smallest eigenvalue is not zero, the graph is declared connected. Finally, the connectivity probability is evaluated over a massive graph ensemble. 

In the literature, other approaches have leveraged graph-theory representations in ad-hoc networks. A common approach is the adjacency-matrix exponent \cite{Hoque2014}, which performs matrix multiplications $k$ times in each iteration to evaluate the connectivity. Both eigenvalue decomposition and matrix multiplication are of the same complexity degree, and therefore, the Eigenvalue method is deemed more computationally efficient as it performs the former once at each iteration. 

\begin{figure}[!t]
    \centering
    \includegraphics[width = 0.7\columnwidth]{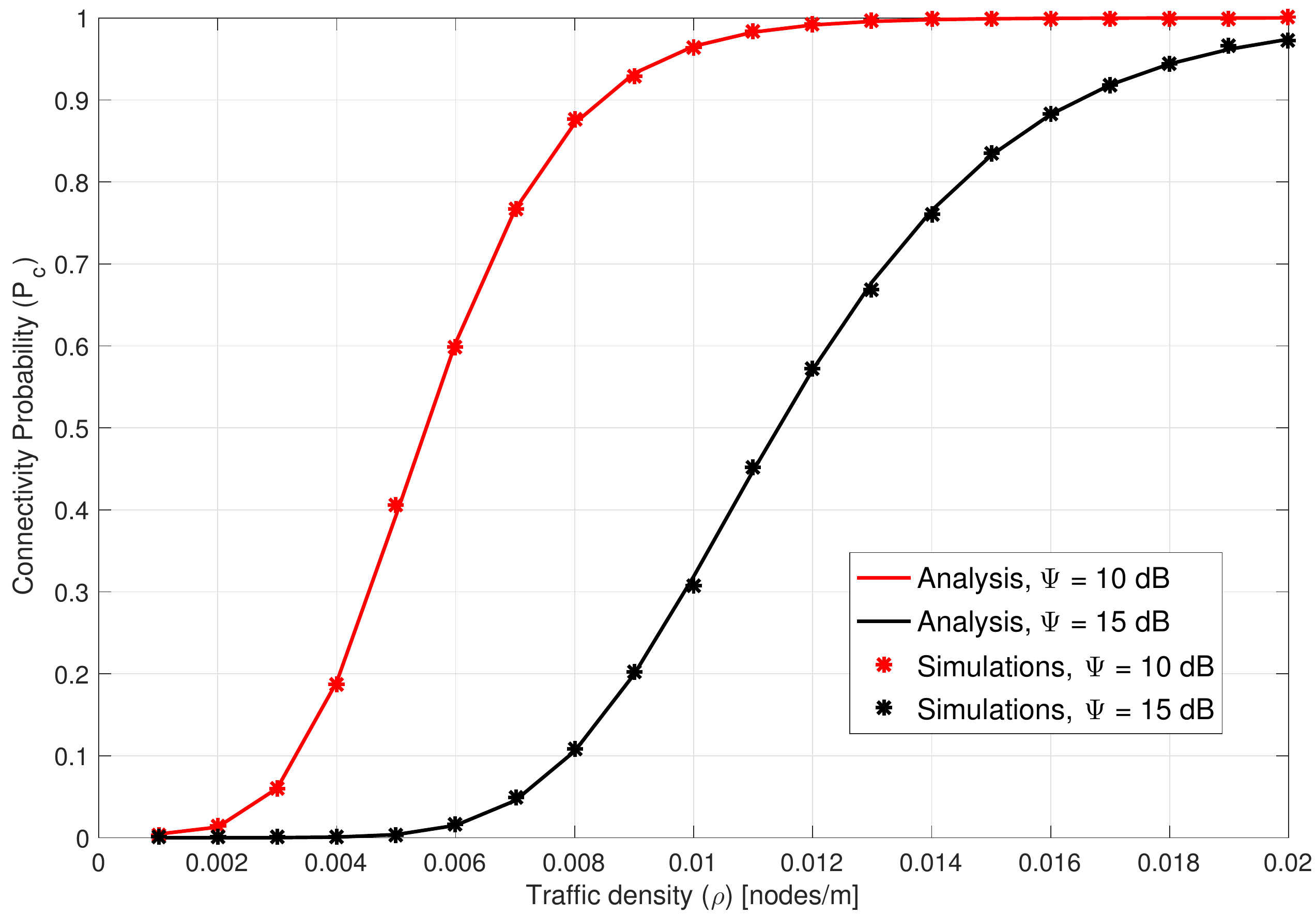}
    \caption{Connectivity probability under a unit disc communication channel ($P_T= 33$ dBm, $P_{noise} = 0.01$ mW, $\beta = 10$, $\alpha = 2$ and $L = 10$ km).}
    \label{fig:FixedCR}
\end{figure}

Simulations results are illustrated in Fig. \ref{fig:FixedCR}. We considered a 10-km multi-lane road segment and repeated the process at different average vehicle densities and different SNR threshold values. The simulation results were found identical to the analytic curves generated from \eqref{Eq:FixedCR_Connectivity}. 
\section{Connectivity in Rayleigh-Fading Channels}\label{Rayleigh}
Wave propagation through real-world communication channels suffers from multipath fading. This, in turn, affects the communication transmissions because the receiving terminals receive a number of delayed versions of the transmitted signals. Under the free-flow traffic, where the vehicles are widely separated, the LoS propagation is almost absent. Therefore, the received signal amplitudes follow a Rayleigh distribution \cite{Acosta-Marum2007}. In this section, we thoroughly study connectivity over a Rayleigh-fading channel. 
\subsection{Intervehicle Distance Distribution}
Over a Rayleigh-fading channel, the vehicles can reach farther neighbors, regardless of their connectivity to their closer neighbors. Therefore, the single-link connectivity should consider any neighbor in the surroundings as first described in Section \ref{Conn}. In this case, the intervehicle distance distribution between any two vehicles in the network should be determined first.

The distance between any pair of vehicles is a random variable that follows a different probabilistic distribution from the exponential distribution, which describes the intervehicle spacing, i.e., the distance between two consecutive vehicles. In order to distinguish between the \textit{intervehicle spacing} and the distance between any two vehicles, we refer to the latter as \textit{intervehicle distance}. Since the intervehicle spacing is exponentially distributed, the intervehicle distance between any vehicle and its $m^{th}$ neighbor is the sum of $m$ independent exponentially-distributed intervehicle spacings (see Fig. \ref{FixedCR_Network}). Let the intervehicle distance to the $m^{th}$ neighbor be $\text{Z}_m$ such that $\text{Z}_m \triangleq \sum_{i=1}^{m} \text{X}_i$, where $\text{X}_i$ with the exponential PDF described in \eqref{PDF_One}. Thus, the PDF of $\text{Z}_m$ would be
\begin{equation}
    f_{\text{Z}_m}(x) = f_{\text{X}_1}(x) * f_{\text{X}_2}(x) * \dots * f_{\text{X}_m}(x)
    \label{n-fold}
\end{equation}
where the operator $*$ is the linear convolution. The solution of the m-fold convolution of exponential distributions follows an Erlang distribution with a PDF of 
\begin{equation}
    f_{\text{Z}_m}(x; m) = \frac{\rho^m}{(m-1)!} x^{m-1} \ e^{-\rho x}, \quad x \geq 0
    \label{intervehicle_Dist}
\end{equation}
where the operator $(.)!$ is the factorial \cite{Akkouchi2008}. The average intervehicle distance is, accordingly, $\mathbb{E}[Z_m;m] = \frac{m}{\rho}$.

\subsection{Single-Link Connectivity}\label{V2V_Connectivity}
While the received signal amplitude is Rayleigh distributed, the received power, and in turn the SNR, follows an exponential distribution \cite{Goldsmith2005}. Let $Y_m$ be a random variable that represents the received SNR by the $m^{th}$ neighbor in a fading channel. The conditional PDF given that the $m^{th}$ neighbor is at a certain distance $x$ is exponentially distributed such that
\begin{equation}
    f_{Y_m|Z_m}(y|x) =  \dfrac{x^{\alpha} P_{noise}}{\beta P_T}\ e^{-\frac{y x^{\alpha} P_{noise}}{\beta P_T}}
    \label{SNR_Z}
\end{equation}
where the conditional average SNR is identical to the received SNR in the unit disc model (cf. \eqref{free-space}) \cite{Miorandi2006} 
\begin{equation}
	\mathbb{E}[Y_m | Z_m = x] = \frac{\beta P_T}{x^{\alpha} P_{noise}}
	\label{Avg_SNR} 
\end{equation}
While the average SNR decreases with the distance, the received SNR is still a random variable. At any given time, the received SNR (i.e., an instance of $Y_m$) at distance $d_2$ could be higher than the SNR at $d_1$ with $d_2 > d_1$. 

In order to find the single-link connectivity to the $m^{th}$ neighbor, we need to find the probability that the received SNR by that neighbor exceeds the threshold $\Psi$. However, in contrast to the unit disc, the approach that checks whether the neighbor is located within the communication range of the vehicle is not adequate in fading channels. When the received SNR at $d_2$ is higher than the one at $d_1$, it means that closer areas does not belong to the transmitter's coverage while farther points do. The definition of the vehicle's communication range is not then well-defined compared to the unit disc. In other words, the equivalence between the communication-range and SNR approaches of evaluating the connectivity fails. Therefore, the analysis and results of the fading-based studies that have relied on this equivalence (e.g., \cite{Babu2013, Chandrasekharamenon2012, C.Neelakantan2012, Miorandi2006, Miorandi2008}) become questionable. It is worth mentioning that these studies concluded that the average communication range in fading channels is lower than the communication range of the unit disc under the same conditions. This suggested that connectivity deteriorates in fading channels.

To proceed in our analysis to find the single-link connectivity, we first derive an expression for the joint PDF of the received SNR and the intervehicle distance from \eqref{SNR_Z}
\begin{equation}
	\begin{split}
		f_{Y_m Z_m}(y,x;m) & = f_{Y_m|Z_m}(y|x) f_{\text{Z}_m}(x; m)\\
		& = \frac{\rho^m P_{noise}}{\beta P_T(m-1)!}x^{\alpha+m-1} \ e^{-\rho x-\frac{y x^{\alpha} P_{noise}}{\beta P_T}}
	\end{split}
	\label{Joint_Ray}
\end{equation}
Then, the marginal PDF of the received SNR at the $m^{th}$ neighbor becomes
\begin{equation}
	f_{Y_m}(y;m)  = \frac{\rho^m P_{noise}}{\beta P_T(m-1)!} \int_0^{\infty}x^{\alpha+m-1} \ e^{-\rho x-\frac{y x^{\alpha} P_{noise}}{\beta P_T}} dx
	\label{Marginal_Ray}
\end{equation}
A closed formula of the average received SNR could be then reached for $m \geq \alpha + 1$ (otherwise, it can be computed numerically);
 \begin{equation}
	\mathbb{E}[Y_m;m]  = \frac{\beta P_T \rho^{\alpha}}{P_{noise}} \prod_{j = 1}^{\alpha} \left( \dfrac{1}{m-j} \right)
	\label{Average_Ray}
\end{equation}

The single-link connectivity can then be derived as the probability that the received SNR exceeds the threshold $\Psi$
\begin{equation}
	\begin{split}
		 P_{SL|Ray} & = \mathbb{P}(Y_m \geq \Psi;m)  = 1 - F_{Y_m}(\Psi;m)\\
		& = \frac{\rho^m P_{noise}}{\beta P_T(m-1)!} \int_{\Psi}^{\infty} \int_0^{\infty}x^{\alpha+m-1} \ e^{-\rho x-\frac{y x^{\alpha} P_{noise}}{\beta P_T}} dx dy \\
		& = \frac{\rho^m}{(m-1)!} \int_0^{\infty}x^{m-1} \ e^{-\rho x-\frac{\Psi x^{\alpha} P_{noise}}{\beta P_T}} dx 
	\end{split}
	\label{SL_conn}
\end{equation}
which should be evaluated numerically. A closed formula of \eqref{SL_conn}, however, can be derived when the PLE is set to $2$;
\begin{equation}
        P_{SL|Ray} =
        \frac{\rho^{m-1}}{(m-1)!}
        \left(\frac{\rho \lambda^2}{2}\right)^m
        \ e^{\frac{\rho^2 \lambda^2}{4}} \sum_{k=0}^{m-1}
        \binom{m-1}{k}
        \left( \frac{-2}{\rho \lambda}\right)^k
        \Gamma\left(\frac{k+1}{2},\frac{\rho^2 \lambda^2}{4}\right)
    \label{V2V_Rayleigh}
\end{equation}
where the $\Gamma(m, x)$ is the upper incomplete gamma function. 

For the sake of comparison, we derive a general form for single-link connectivity in a unit disc scenario, where any neighbor is considered, as follows
\begin{equation}
    \begin{split}
        P_{SL|UD} & = \mathbb{P}(\text{Z}_m \leq r; m)\\
        & = \frac{\rho^m}{(m-1)!} \int_{0}^{r} x^{m-1} \ e^{-\rho x}dx
        = 1 - e^{-\rho r} \sum_{k = 0}^{m - 1} \frac{(\rho r)^k}{k!}
    \end{split}
    \label{V2V-Fixed}
\end{equation}
where $r$ is the fixed communication range. It is clearly obvious that the single-link connectivity to the closest neighbor, i.e., $m = 1$, tends to \eqref{exponential_Conn}. 

\begin{figure}
	\centering
        \includegraphics[width = 0.7\columnwidth]{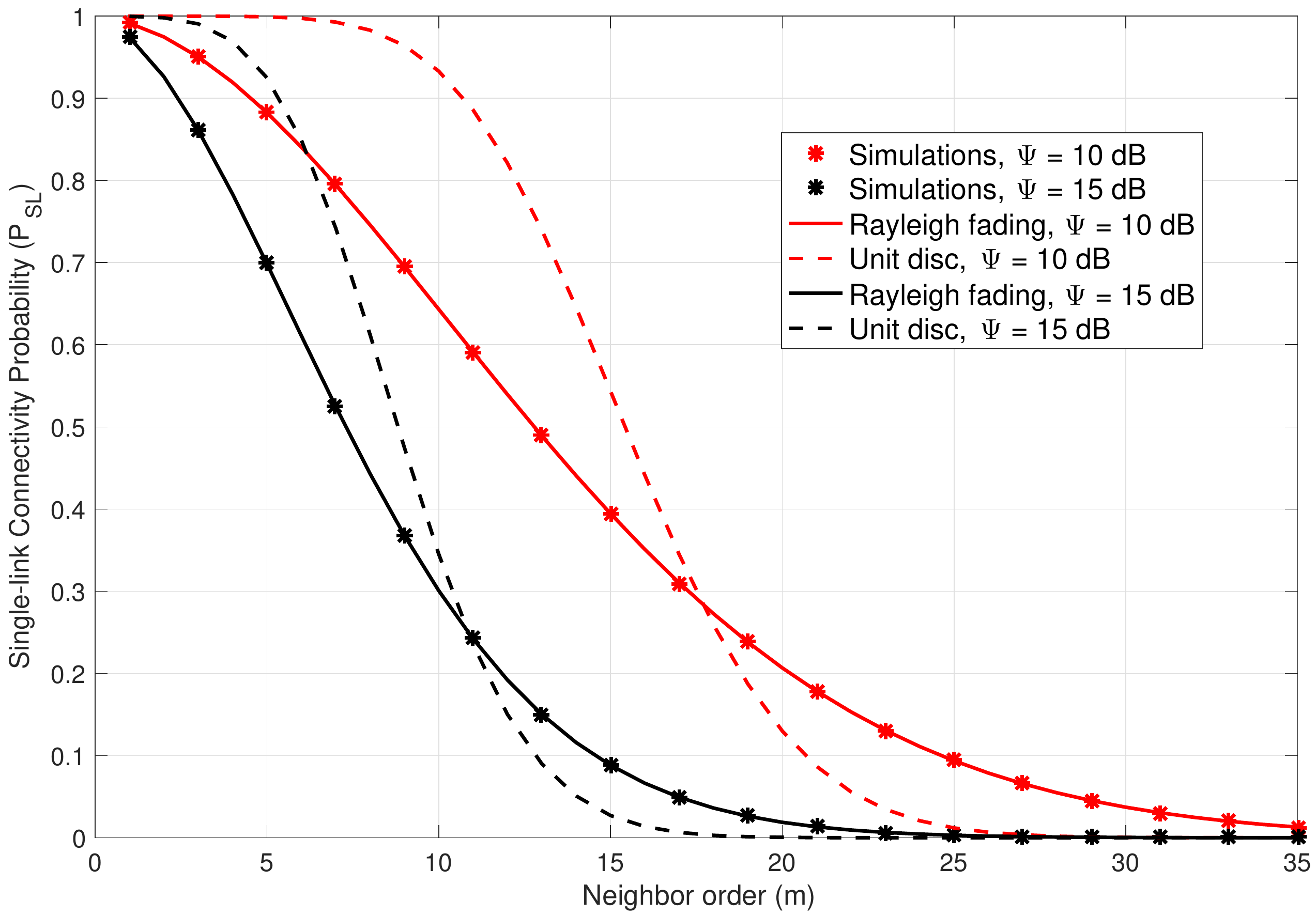}
    \caption{Single-link connectivity probability to different neighbors at two different SNR thresholds ($\rho = 0.019$ vehicles/m, $P_T = 33$ dBm, $P_{noise} = 0.01$ mW, $\beta = 10$, $\alpha = 2$, and $L = 10$ km).}
    \label{fig:V2V}
\end{figure}
The single-link connectivity of a Rayleigh-fading channel is represented as solid lines in Fig. \ref{fig:V2V}. As depicted, the single-link connectivity to the closest neighbors falls behind that of the unit disc. Surprisingly, that behavior was completely reversed for the farthest neighbors. Long links, i.e., connections to further neighbors, have higher probabilities to occur due to fading. Vehicles are even able to reach new neighbors further than the unit disc coverage. This remark matches the results reported in \cite{Bettstetter2005, Hekmat2006, Zhou2008, Georgiou2014} for other ad-hoc networks. 

Graph-based simulations were also implemented to validate the theoretical results. To perform these simulations, Algorithm \ref{algo} was modified as follows. The required information about the single-link connectivity was gathered from the off-diagonal elements of the adjacency matrix $\mathbf{A}$ in Step 9, and there was no need to proceed to the following steps. Besides, the received SNR in Step 8 was chosen to be an exponential random variable to mimic the Rayleigh fading channel in equations \eqref{SNR_Z} and \eqref{Avg_SNR}. The simulation results matched the theoretical ones as illustrated in Fig. \ref{fig:V2V}.
\subsection{Average Node Degree} \label{Node degree}
\begin{figure}[!t]
    \centering
    \includegraphics[width = 0.7\columnwidth]{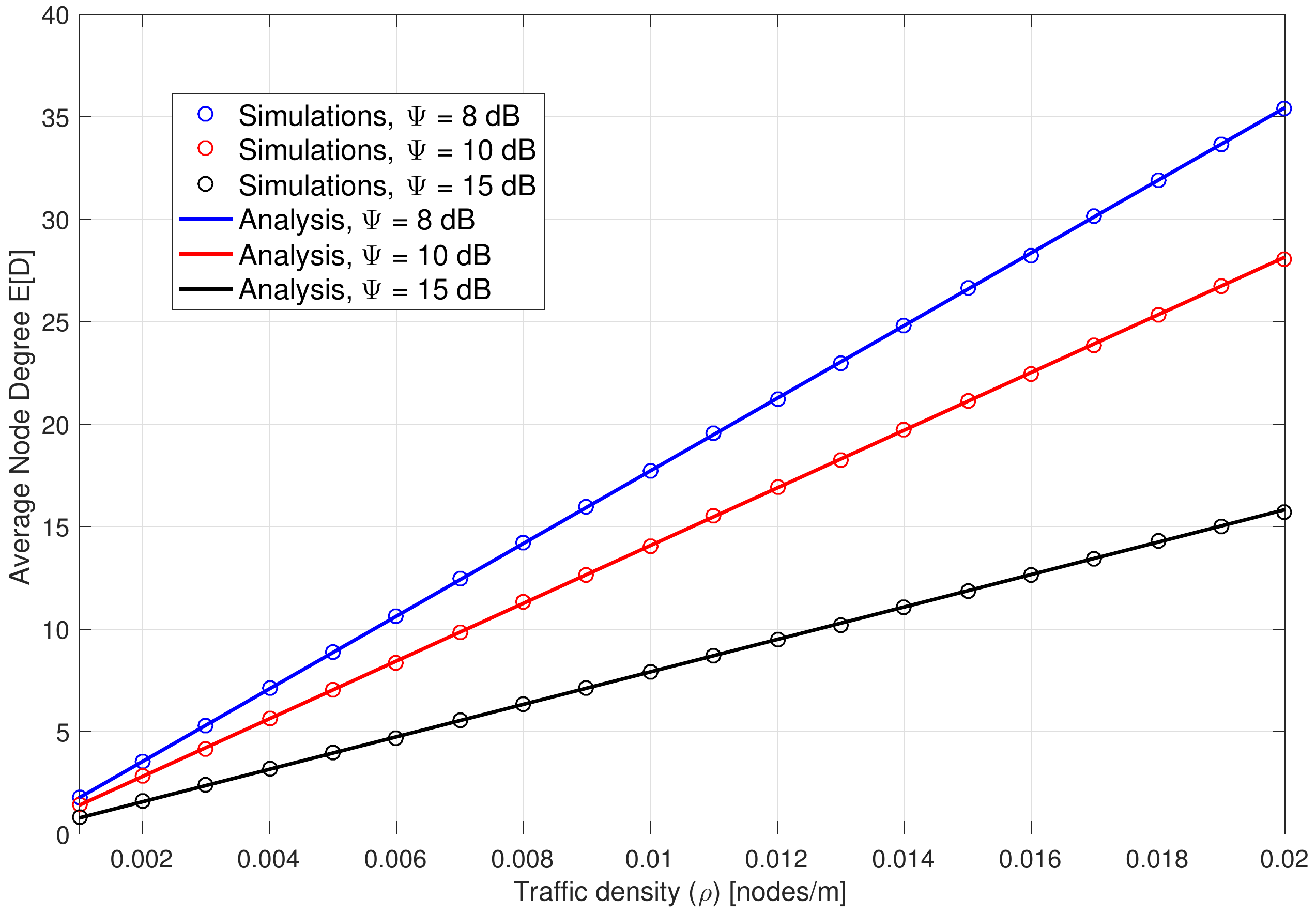}
    \caption{Average node degree of Rayleigh-fading channels ($P_T = 33$ dBm, $P_{noise} = 0.01$ mW, $\beta = 10$, $\alpha = 2$, and $L = 10$ km).}
    \label{fig:NodeDeg}
\end{figure}

The average node degree is the average number of linked neighbors of every vehicle in the network. As a generalization of \eqref{SNR_Z}, one could consider the conditional PDF of the received SNR at any distance $x$ to be 
\begin{equation}
    f_{Y|X}(y|x) =  \dfrac{x^{\alpha} P_{noise}}{\beta P_T}\ e^{-\frac{y x^{\alpha} P_{noise}}{\beta P_T}}
    \label{SNR_X}
\end{equation}
where $X$ is a random variable that reperesents the distance from the transmitting vehicle. Hence, the probability that the received SNR at a distant $x$ exceeds a threshold $\Psi$ is
\begin{equation}
	\begin{split}
		\mathbb{P}(Y \geq \Psi|X = x) & = \frac{P_{noise}}{\beta P_T} \int_{\Psi}^{\infty} x^{\alpha} \ e^{- \frac{y x^{\alpha} P_{noise}}{\beta P_T}} dy \\
		& = e^{- \frac{x^{\alpha} \Psi P_{noise}}{\beta P_T}} 
	\end{split}
\end{equation}
The average node degree can then be determined as \cite{Bettstetter2005}
\begin{equation}
	\mathbb{E}[D] = \rho \int_{-\infty}^{\infty} e^{- \frac{x^{\alpha} \Psi P_{noise}}{\beta P_T}}  dx
\end{equation}
which is verified with simulations in Fig. \ref{fig:NodeDeg}.

In the unit disc model, a vehicle can be connected simultaneously to a number of vehicles, as implied from Fig. \ref{fig:V2V}. However, only the connection to the first successor is needed to keep the network connected, and the others are dependent connections as we discussed in Section \ref{Model}. Therefore, a VANET under the unit disc assumption is sufficiently represented by an undirected graph with a node degree that is lower than or equals 2.  The difference in the node degree implies that the Rayleigh-fading model captures the possibility of transmitting the packets through different routes. If one of those routes is dropped because of fading, the transmission might succeed through another route. However, the unit disc assumption embraces one independent route. If it is dropped, the unit disc neglects any other possibility of maintaining the connectivity.
\subsection{Vehicle Connectivity} \label{VehicleConnect}
\begin{figure}[!t]
    \centering
    \begin{subfigure}[b]{0.7\columnwidth}
        \includegraphics[width = \columnwidth]{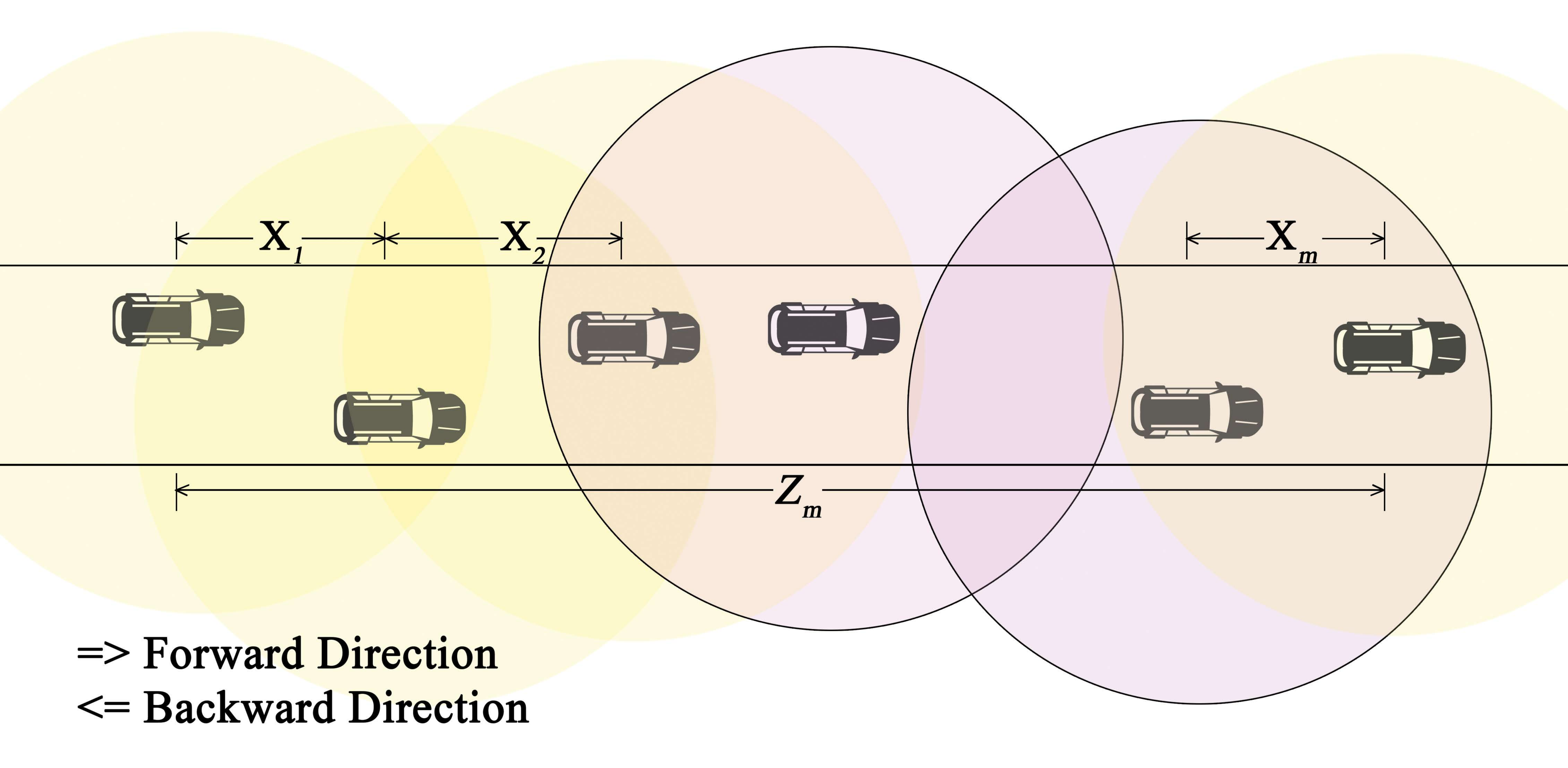}
        \caption{}
        \label{FixedCR_Network}
    \end{subfigure}
    \begin{subfigure}[b]{0.7\columnwidth}
        \includegraphics[width = \columnwidth]{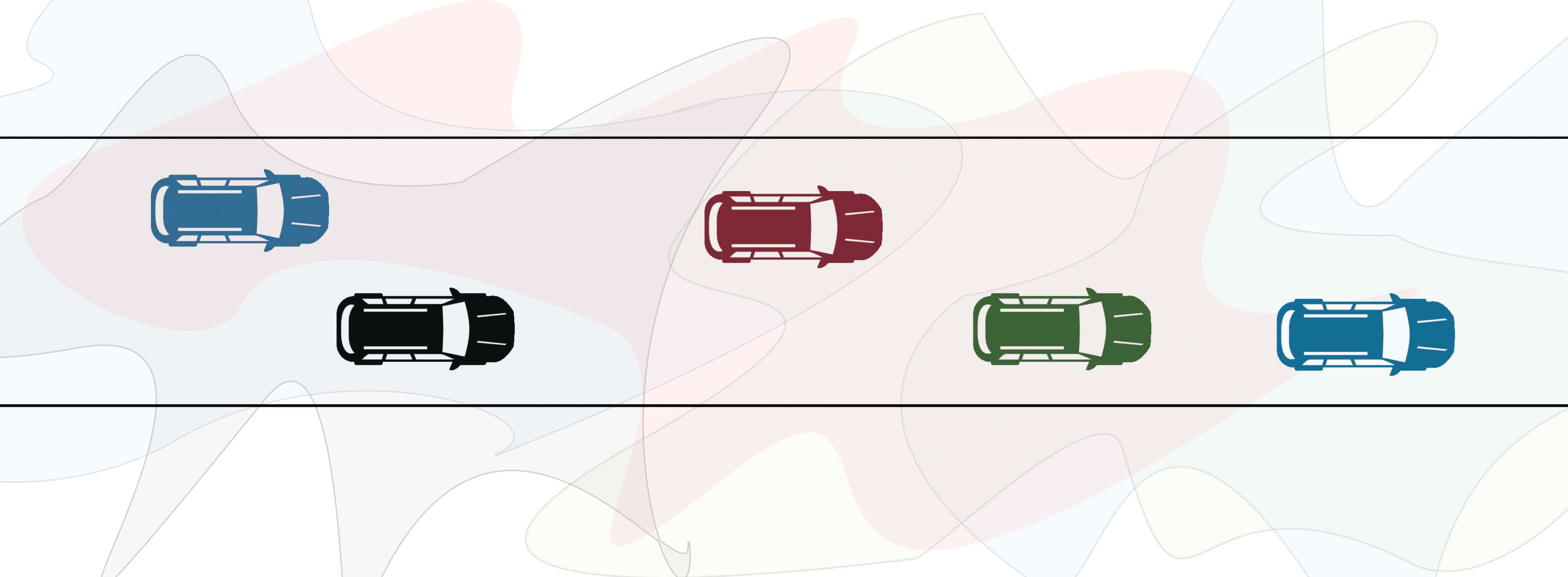}
        \caption{}
        \label{Rayleigh_Network}
    \end{subfigure}
    \caption{An illustration of a VANET showing the effect of having only one vehicle that is one-side isolated in two different communication channel models. (a) A unit disc channel, where the third (second) vehicle from the right is one-side isolated when the forward (backward) direction is considered, leading to preventing the network connectivity. (b) A fading channel model with randomly-distributed communication ranges. The black vehicle is isolated from the forward direction but it does not prevent the network connectivity.}
    \label{fig:Network}
\end{figure}

We define vehicle isolation according to the sufficient conditions required by a vehicle to eliminate the network connectivity. Thereafter, we can evaluate how a single vehicle affects the network connectivity in the two different channel models. Accordingly, we get two different definitions of vehicle isolation, each of which matches one of the considered channel models. In a unit disc, a blocked connectivity to one side of a vehicle is sufficient and necessary to drop the connectivity. By a one-side vehicle connectivity, we only consider the connections in a predefined direction, either the forward or the backward direction. In Fig. \ref{FixedCR_Network}, if the forward direction is considered, the third vehicle from the right is disconnected from its forward neighbors, leading to preventing the network connectivity. The same result can be concluded when the backward direction is considered; the second vehicle from the right is, then, the isolated node that blocks the network connectivity. Thus, the vehicle connectivity in the unit disc channels can be narrowed to be one-sided. 

In contrast, the one-side isolation is not sufficient to prevent connectivity in Rayleigh-fading environments. As in Fig. \ref{Rayleigh_Network}, although the coverage of the black vehicle does not contain any forward neighbors, its connectivity is maintained by its link to its first backward neighbor. Consequently, the two-side isolation is sufficient but not necessary for burdening the connectivity, and then it is the convenient candidate for describing the vehicle connectivity under the Rayleigh-fading assumptions.

In general, the probability that a vehicle is connected from one side can be expressed as
\begin{equation}
        P_{1} = \mathbb{P} \left( \bigcup_{m = 1}^{M} (Y_m \geq \Psi; m)\right)
        = 1-  \mathbb{P} \left( \bigcap_{m = 1}^{M} (Y_m < \Psi; m)\right)
    \label{vehicle_connectivity}
\end{equation}
where $M$ should be the whole number of nodes located on one side of a vehicle. However, this number can be reduced to span only the neighbors within the vehicle's proximity in order to relax the computations. 

With a Rayleigh-fading channel, the single-link connectivity to different neighbors can be reasonably viewed as uncorrelated but dependent events. As the connectivity is controlled by the communication channel to different neighbors, which can be assumed uncorrelated in a Rayleigh channel if the received antennas are at least spaced by a wavelength apart. Besides, it also depends on the distance to different nonconsecutive neighbors $\text{Z}_m$, which is relatively dependent since $\text{Z}_{m+1}\geq \text{Z}_m$ for $m = 1, \dots M$. Consequently, the vehicle connectivity should be expressed in terms of conditional probabilities. For simplicity, we  assume that the single-link connectivity to different neighbors are independent of each other in order to derive an approximation of the vehicle connectivity. Thus, the vehicle connectivity from one side becomes
\begin{equation}
	\begin{split}
        		P_{1|Ray}  & = 1 - \prod_{m = 1}^{M}  \mathbb{P}(Y_m < \Psi; m)\\
		& = 1 - \prod_{m = 1}^{M} \left( 1 - \frac{\rho^m}{(m-1)!} \int_{0}^{\infty} x^{m-1} \ e^{-\rho x - \left( \frac{x}{\lambda}\right)^\alpha}dx \right)
        \end{split}
    \label{One-side}
\end{equation}
Since the vehicle disconnectivity from one side is independent of its disconnectivity from the other side, the vehicle connectivity in a Rayleigh-fading channel can be evaluated as 
\begin{equation}
    \begin{split}
        P_{V|Ray} & = 1 - \left( \prod_{m = 1}^{M}  \mathbb{P}(Y_m < \Psi; m) \right)^2\\
        & = 1 - \left( \prod_{m = 1}^{M} \left( 1 - \frac{\rho^m}{(m-1)!} \int_{0}^{\infty} x^{m-1} \ e^{-\rho x - \left( \frac{x}{\lambda}\right)^\alpha}dx \right) \right)^2
    \end{split}
    \label{Two-side}
\end{equation}

On the other hand, the vehicle connectivity of a unit disc is much simpler as the connections to remote neighbors do not represent unique situations. 
Therefore, the vehicle connectivity in \eqref{vehicle_connectivity} tends to be the single-link connectivity probability to the first neighbor, which can be declared as
\begin{equation}
        P_{V|UD} = \mathbb{P}(Z_m \leq r; m = 1) = 1 - \ e^{-\rho r}
\end{equation}

\begin{figure}[!t]
    \centering
    \includegraphics[width = 0.7\columnwidth]{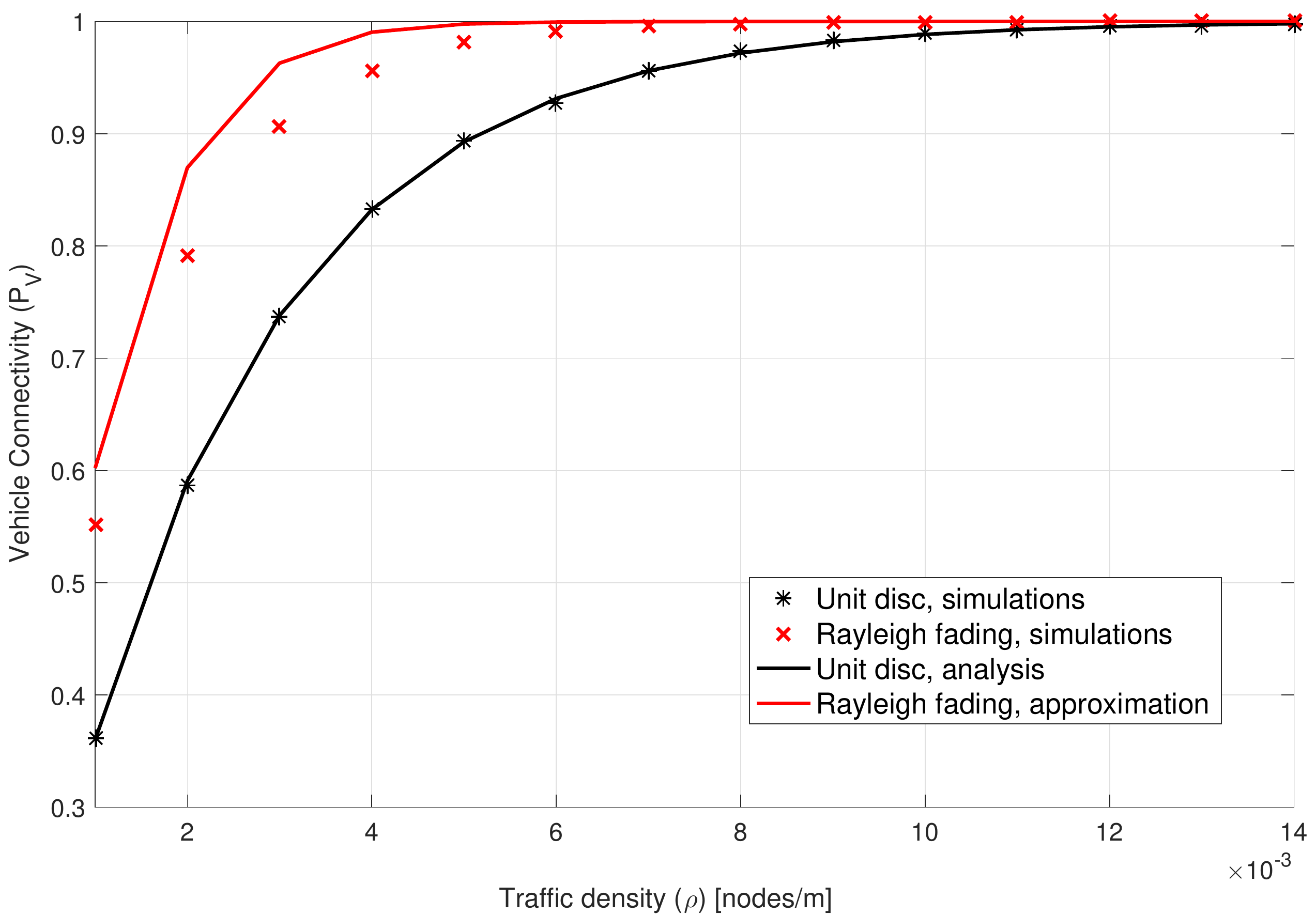}
    \caption{Vehicle connectivity probability for $\Psi = 15$ dB, ($P_T = 33$ dBm, $P_{noise} = 0.01$ mW, $\beta = 10$, $\alpha = 2$, $M = 10$, and $L = 10$ km).}
    \label{fig:Pv}
\end{figure}

Graph-based simulations were executed using the proposed algorithm in Section \ref{Graph}. The vehicle connectivity is related to and can be evaluated from the degree matrix $\mathbf{D}$ in Step 10 of Algorithm \ref{algo}. A comparison between vehicle connectivity in both Rayleigh-fading and unit disc channels based on the simulation results is represented in Fig. \ref{fig:Pv}.
As shown in the figure, the (two-side) vehicle isolation, i.e., the complement of vehicle connectivity is lower than the (one-side) vehicle isolation of the unit disc. Networks at fade preserve connectivity through variant network topologies, which gives each vehicle multiple ways to stay connected. However, the unit disc model assumes only one topology, which reduces the vehicle connectivity probability. Fig. \ref{fig:Pv} also shows that the vehicle connectivity of \eqref{Two-side} serves as an upper bound for the actual vehicle connectivity. The discrepancy is resulted from tolerating the dependence between the intervehicle distances.
\section{Graph-based Simulations of the Improved Connectivity}\label{Improvement}
With improved vehicle connectivity under a Rayleigh-fading channel, it is crucial to analyze the VANET connectivity to detect the fading implications on it. The definition of connectivity in unit disc channels cannot be extended to fading channels. In the unit disc, it is sufficient and necessary to check whether each vehicle is connected to its successor, the property that is not necessary in a fading channel. The connectivity can be retained by many other network topologies. Furthermore, even if each vehicle is not isolated, the whole network may still be partitioned. That leaves defining a closed formula for the connectivity probability nearly infeasible. Therefore, we rely on graph theory to numerically predict the connectivity probability in fading channels with Monte-Carlo simulations.

\begin{figure}[!t]
    \centering
    \includegraphics[width = 0.7\columnwidth]{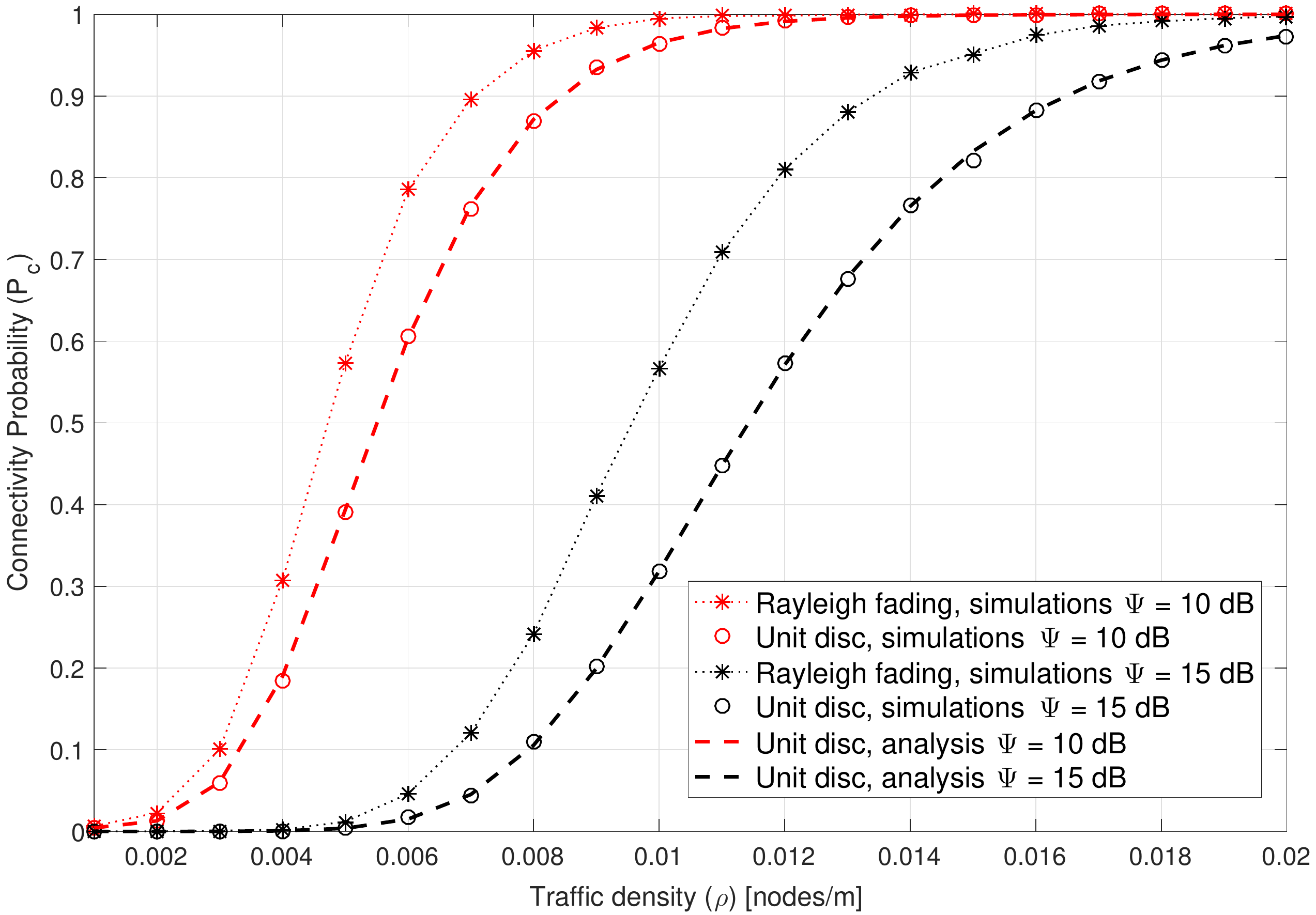}
    \caption{Improved connectivity probability of Rayleigh-fading channels over the unit disc ($P_T = 33$ dBm, $P_{noise} = 0.01$ mW, $\beta = 10$, $\alpha = 2$, and $L = 10$ km).}
    \label{fig:Connectivity}
\end{figure}

The graph Monte-Carlo simulations discussed in Section \ref{Graph} provide an exact connectivity probability by averaging the number of connected graphs in a $10^3$-graph ensemble. The graph ensemble represents VANET over a road segment of length $L = 10$ km. Algebraic connectivity determines the network connectivity, and therefore, the utilized simulations are independent of the probabilistic analysis. The comparison between the unit disc and Rayleigh-fading models hold by fixing the threshold $\Psi$ in the two cases. In the simulations, we fixed the values of $P_T = 33$ dBm and $P_{noise} = 0.01$ mW, $\beta = 10$ and the PLE was  $\alpha = 2$. The simulations were executed at different vehicle densities, and the results are shown in Fig. \ref{fig:Connectivity}. 

The simulation results show the improved connectivity of Rayleigh-fading channels over the unit disc. For different threshold values, Rayleigh channels achieve higher connectivity probability $P_c$. The connectivity curves tend to reach the unity probability faster in the Rayleigh-fading channel, providing higher connectivity probabilities at lower vehicle densities. It is also noticeable that the difference between the two models increases at a higher SNR threshold, where the disk's radius shrinks. In other words, when lower communications ranges are considered, the unit-disc coverage becomes very limited and is not comparable to the variety of longs links provided in a Rayleigh channel. Therefore, the average behavior of VANETs captured by Monte-Carlo simulations negates the claims that the unit disk shares the same manners. The motives of these claims about the unit disc are, however, discussed in Appendix \ref{app}.

The aforementioned advantages were accomplished because of the reinforced connections to further vehicles. This, in turn, leads to connecting more separated partitions of the network and raising the connectivity probability. Besides, the variety of available connected topologies provides the network with \textit{diversity}. If the links to some of them experience deep fade, others can mitigate it and provide connectivity. Although other types of diversity are manmade, the diversity provided here is granted by the cooperation manner of ad-hoc networks. 

\section{Conclusions} \label{Conclusions}
Rayleigh-fading models were shown to have an improved connectivity compared to the unit disc model. Due to the lack of necessary and sufficient conditions for the network connectivity at fading, it is nearly infeasible to describe this behavior with closed formulas. Graph-based Monte-Carlo simulations were the most convenient method to capture this behavior of Rayleigh channels. Although the concluded remark  in this paper seems counterintuitive, it was justified by the long links to further neighbors that cannot be caught by the unit disk model. Along with the diversity acquired by the variety of connected network topologies, fading helps to boost the network connectivity. 
\appendices
\section{The Agreement on the Unit Disc Model}\label{app}
The wide agreement on the unit disc model as a representative model of the average behavior of VANETs \cite{Naboulsi2017} can be justified by the analogy between the average SNR provided in both cases. The average SNR received by the $m^{th}$ neighbor in the unit disc is identical to the one of the Rayleigh-fading channels. This can be proved following the same analysis as in equations \eqref{Joint_Ray}-\eqref{Average_Ray} starting with the conditional PDF of the SNR given the intervehicle distance
\begin{equation}
	\begin{split}
		f_{\Lambda_m | Z_m}(y|x) & = 
		\begin{cases}
			1 & y = \frac{\beta P_t}{x^{\alpha} P_{noise}}\\
			0 & otherwise
		\end{cases}
		\\
		& =  \delta \left( y - \dfrac{\beta P_t}{x^{\alpha} P_{noise}} \right) 
	\end{split}
\end{equation}
where $\Lambda_m$ is a random variable that represents the received SNR by the $m^{th}$ neighbor under the unit disc assumptions. The marginal PDF of the received SNR has the form
\begin{equation}
f_{\Lambda_m}(y)= \frac{\rho^m}{(m-1)!} \int_0^{\infty} x^{m-1} \ e^{-\rho x} \delta \left( y - \dfrac{\beta P_t}{x^{\alpha} P_{noise}} \right) dx 
\end{equation}
Accordingly, the average SNR received by the $m^{th}$ neighbor (for $m \geq \alpha + 1$) is
\begin{equation}
	\mathbb{E}[\Lambda_m;m]  = \frac{\beta P_T \rho^{\alpha}}{P_{noise}} \prod_{j = 1}^{\alpha} \left( \dfrac{1}{m-j} \right)
	\label{Average_unit}
\end{equation}
which is identical to the average SNR in a Rayleigh-fading channel (see  \eqref{Average_Ray}). 

Despite the mach between the average values, this claim ignores the difference in the PDF of the received SNR under the two cases.  The PDF discrepancy has resulted into divergence in the single-link connectivity of the unit disc from the one in Rayleigh-fading, shown in Fig. \ref{fig:V2V}. Accordingly, all other connectivity metrics we had examined showed superiority in connectivity of the Rayleigh-fading model over the unit disc. 
\end{document}